\documentclass[12pt]{article}
\usepackage{natbib,graphicx,setspace,amsmath,amssymb,subfigure,url}
\newtheorem{thm}{Theorem}

\newtheorem{prp}{Proposition}

\begin{document}

\title{A simple and efficient algorithm for fused lasso signal approximator with convex loss function}         
\author{Heng Lian\\Division of Mathematical Sciences\\School of Physical and Mathematical Sciences\\Nanyang Technological University\\Singapore 637371\\Singapore\\E-mail: henglian@ntu.edu.sg}        
\date{}          
\maketitle
\begin{abstract}
We consider the augmented Lagrangian method (ALM) as a solver for the fused lasso signal approximator (FLSA) problem. The ALM is a dual method in which squares of the constraint functions are added as penalties to the Lagrangian. In order to apply this method to FLSA, two types of auxiliary variables are introduced to transform the original unconstrained minimization problem into a linearly constrained minimization problem.  Each updating in this iterative algorithm consists of just a simple one-dimensional convex programming problem, with closed form solution in many cases. While the existing literature mostly focused on the quadratic loss function, our algorithm can be easily implemented for general convex loss. The most attractive feature of this algorithm is its simplicity in implementation compared to other existing fast solvers. We also provide some convergence analysis of the algorithm. Finally, the method is illustrated with some simulation datasets.

\textbf{keywords:} Augmented Lagrangian; Convergence analysis; LAD-FLASSO;
\end{abstract}

\section{Introduction}       
In this paper we examine the one-dimensional fussed lasso signal approximator \citep{tibshirani05}, which is to solve
\begin{equation}\label{eqn:flsa}
\min_\beta f(\beta)=F(y,\beta)+\lambda_1\sum_{i=1}^n|\beta_i|+\lambda_2\sum_{i=2}^n|\beta_i-\beta_{i-1}|,
\end{equation}
where $y=(y_1,\ldots,y_n)$ are the noisy observations, $\lambda_1,\lambda_2>0$ are two regularization parameters and $F(y,\beta)=\sum_{i=1}^nF_i(\beta_i,y_i)$ is the loss function. The most frequently appearing case is the quadratic loss $F_i(\beta_i,y_i)=(y_i-\beta_i)^2/2$, for which there exists several solvers. Here we also consider the more general case where $F_i$ is a convex and coercive function of $\beta_i$. Note that by definition the coercive function $F_i$ satisfies $\lim_{|\beta_i|\rightarrow\infty} F_i(\beta_i,y_i)\rightarrow\infty$, for all $y_i\in R$, which is used to ensure the existence of the minimizer. As demonstrated in \cite{huang05,tibshirani08}, an important application of FLSA is the reconstruction of copy numbers from CGH arrays. 

Several algorithms have been proposed for FLSA, including a specially designed quadratic programming \citep{tibshirani05,tibshirani08}, coordinate descent and fusion algorithm \citep{friedman07} and a path algorithm that solves the problem for all regularization parameters simultaneously \citep{hoefling10}. Based on the numerical results performed in \cite{hoefling10}, the latter two algorithms are clearly very fast and efficient and represent the state of the art. However, these two algorithms require substantial efforts in implementation for non-expert programmers, since one needs to keep track of the ``fused sets" which contains the coefficients $\beta_i$ that assume the same value. Besides, the algorithm of \cite{friedman07} has the disadvantage that once the coefficients are fused, the linkage cannot be removed later (a similar problem is noticed in \cite{zou08} for the locally quadratic approximation algorithm proposed in variable selection problem with non-concave penalty), and no convergence analysis is available. The algorithm of \cite{hoefling10} is designed to solve (\ref{eqn:flsa}) for all regularization parameters but it does not work for general convex loss since the solution path is in general not piecewise linear \citep{rosset07}. 

Here we consider augmented Lagrangian method (ALM) which was independently developed by \cite{hestenes69} and \cite{powell69} almost half a century ago, which aims to solve convex optimization problem with linear constraints. There are surged interests recently in applying this method in different optimization problems \citep{tai09,tao10,wen09,yang10,yang10b}. We will show that after some simple transformations of (\ref{eqn:flsa}), the ALM can be applied to efficiently solve FLSA with general loss functions. The most attractive  feature of the method is its simplicity in implementation. We present our R code for solving (\ref{eqn:flsa}) with quadratic loss in Appendix B in the Supplementary Material, in which the main iterations consist of only about 20 lines of commands. We provide a clear self-contained convergence analysis of ALM in our context (Appendix A in the Supplementary Material) following existing ideas. Our algorithm can be initialized essentially arbitrarily, in particular initialized with zero values, while for algorithms of \cite{friedman07,hoefling10} such initialization will not work and the coefficients will stay at zero at all times. 

\section{Augmented Lagrangian Formulation}
By introducing the auxiliary variables $\theta_i, i=2,\ldots, n$, the following linearly constrained problem is trivially equivalent to (\ref{eqn:flsa}).
\begin{eqnarray*}
\min_{\beta,\theta}&& g(\beta,\theta)=\sum_{i=1}^nF_i(y_i,\beta_i)+\lambda_1\sum_{i=1}^n|\beta_i|+\lambda_2\sum_{i=2}^n|\theta_i|\\
s.t. &&\theta_i=\beta_i-\beta_{i-1}, i=2,\ldots,n.
\end{eqnarray*}
Following \cite{glowinski89}, we define the augmented Lagrangian, for $c>0$, by
\begin{equation*}
\mathcal{L}_c(\beta,\theta,\nu)=g(\beta,\theta)+\sum_{i=2}^n\nu_i(\theta_i-\beta_i+\beta_{i-1})+\frac{c}{2}\sum_{i=2}^n(\theta_i-\beta_i+\beta_{i-1})^2,
\end{equation*}
where $\nu=(\nu_2,\ldots,\nu_n)$ is the Lagrange multiplier.

We consider the following saddle-point problem,
\begin{eqnarray}
\mbox{ Find } &&\beta^*,\theta^*,\nu^*, \nonumber\\
s.t. && \mathcal{L}_c(\beta^*,\theta^*,\nu)\le \mathcal{L}_c(\beta^*,\theta^*,\nu^*)\le \mathcal{L}_c(\beta,\theta,\nu^*), \;\forall\; \beta,\theta,\nu.\label{eqn:saddle}
\end{eqnarray}
The proof for the following is well known from classical duality theory \citep{rockafellar70,ekeland83} and is thus omitted.
\begin{prp}
$\beta^*$ is a solution of (\ref{eqn:flsa}) if and only if $(\beta^*,\theta^*,\nu^*)$ is a solution of (\ref{eqn:saddle}) for some $\theta^*$ and $\nu^*$.
\end{prp}

The basic algorithm for finding the saddle point is the following Algorithm 1 \citep{glowinski89}.

\bigskip

\begin{tabular}{|l|}
\hline
Algorithm 1\\
\hline
initialize $\nu^0$, arbitrarily.\\
For $k=1,2,\ldots$\\
$\;\;\;\;  (\beta^k,\theta^k)=\arg\min_{(\beta,\theta)} \mathcal{L}_c(\beta,\theta,\nu^{k-1})$\\
$\;\;\;\;\nu^{k}_i=\nu^{k-1}_i+c(\theta_i^k-\beta_i^k+\beta_{i-1}^k),i=2,\ldots,n$\\
\hline
\end{tabular}

\bigskip

In general, it is difficult to minimize $\mathcal{L}_c(\beta,\theta,\nu^k)$ over $\beta$ and $\theta$ simultaneously, but it might be easier to minimize over $\beta$ when fixing $\theta$ and vice versa. In this case, we can alternate these two steps until convergence. It turns out that we can update $\beta$ and $\theta$ just once when the other is fixed, resulting in the following algorithm. 

\bigskip

\begin{tabular}{|l|}
\hline
Algorithm 2\\
\hline
initialize $\nu^0$ and $\theta^0$, arbitrarily.\\
For $k=1,2,\ldots$\\
$\;\;\;\;  \beta^k=\arg\min_{\beta} \mathcal{L}_c(\beta,\theta^{k-1},\nu^{k-1})$\\
$\;\;\;\;  \theta^k=\arg\min_{\theta} \mathcal{L}_c(\beta^{k},\theta,\nu^{k-1})$\\
$\;\;\;\;  \nu^{k}_i=\nu^{k-1}_i+c(\theta_i^k-\beta_i^k+\beta_{i-1}^k),i=2,\ldots,n$\\
\hline
\end{tabular}

\bigskip

\textit{Example.} We apply Algorithm 2 to (\ref{eqn:flsa}) with quadratic loss. In this case, the augmented Lagrangian is 
\begin{equation*}
\mathcal{L}_c(\beta,\theta,\nu)=\frac{1}{2}\sum_i(y_i-\beta_i)^2+\lambda_1\sum_{i=1}^n|\beta_i|+\lambda_2\sum_{i=2}^n|\theta_i|+\sum_{i=2}^n\nu_i(\theta_i-\beta_i+\beta_{i-1})+\frac{c}{2}\sum_{i=2}^n(\theta_i-\beta_i+\beta_{i-1})^2.
\end{equation*}
If $\lambda_1=0$, given $\theta^{k-1}$ and $\nu^{k-1}$, the minimization over $\beta$ is a simple quadratic problem and all components of $\beta$ can be found simultaneously by solving a linear system $B\beta=b$, where we do not write down explicitly the expression of matrix $B$ and vector $b$, but note that due to the special structure of the problem, $B$ is a tridiagonal matrix and there exists an efficient algorithm with complexity linear in $n$ for solving the linear system (see for example \cite{conte80}). 

For $\lambda_1>0$, it is more difficult to update $\beta$ directly. Fortunately, for quadratic loss, solution for FLSA with $\lambda_1>0$ can be obtained by thresholding the solution for FLSA with $\lambda_1=0$ as shown in \cite{friedman07}, and thus (for this example) we only consider $\lambda_1=0$.

With $\beta=\beta^k$ and $\nu=\nu^{k-1}$ fixed, the minimization over $\theta$ is a simple lasso regression with orthogonal design and thus we have the simple component-wise soft thresholding updating rule
\begin{equation}\label{eqn:updatetheta}
\theta_i^k=sign(\hat{\theta}_i)(|\hat{\theta}_i|-\lambda_2/c)_+\;,
\end{equation}
 where $\hat{\theta}_i=\beta_i^k-\beta_{i-1}^k-\nu_i^{k-1}/c$ and $(a)_+$ denotes the positive part of $a$. $\Box$

For quadratic loss, the example shows that both update for $\beta$ and for $\theta$ can be computed efficiently for $\lambda_1=0$. However, for more general loss and/or for $\lambda_1>0$, it is difficult to update $\beta$ directly and thus in our implementation we do not use Algorithms 1 and 2. We propose next a further augmentation step that decouples the quadratic term $(\theta_i-\beta_i+\beta_{i-1})^2$ with the loss function.

We introduce another set of auxiliary variables $\gamma_i, i=1,\ldots, n$ and consider the following problem which is still obviously equivalent to (\ref{eqn:flsa}).
\begin{eqnarray*}
\min_{\gamma,\beta,\theta}&& g(\gamma,\beta,\theta)=\sum_{i=1}^nF_i(y_i,\gamma_i)+\lambda_1\sum_{i=1}^n|\gamma_i|+\lambda_2\sum_{i=2}^n|\theta_i|\\
s.t. &&\gamma_i=\beta_i, i=1,\ldots,n, \;\theta_j=\beta_j-\beta_{j-1}, j=2,\ldots,n.
\end{eqnarray*}

The corresponding (doubly) augmented Lagrangian is 
\begin{eqnarray}
\mathcal{L}_c(\gamma,\beta,\theta,\mu,\nu)&=&g(\gamma,\beta,\theta)+\sum_{i=1}^n\mu_i(\gamma_i-\beta_i)+\frac{c}{2}\sum_{i=1}^n(\gamma_i-\beta_i)^2 \label{eqn:lagrangian2}\\
&&+\sum_{i=2}^n\nu_i(\theta_i-\beta_i+\beta_{i-1})+\frac{c}{2}\sum_{i=2}^n(\theta_i-\beta_i+\beta_{i-1})^2.\nonumber
\end{eqnarray}
In the above, the coefficients for both quadratic penalties are the same (equal to $c/2$). In principle, we can use different coefficients but computationally it is difficult to tune both parameters and thus we settle with this simpler choice. 

With the newly defined Lagrangian in (\ref{eqn:lagrangian2}), we can similarly modify the saddle-point problem (\ref{eqn:saddle}) in an obvious way and it can be shown that the saddle-point problem is the same as the original FLSA problem (\ref{eqn:flsa}). Accordingly, we have the following algorithms for finding the saddle point which directly extends Algorithm 1 and Algorithm 2 respectively.

\bigskip

\begin{tabular}{|l|}
\hline
Algorithm 3\\
\hline
initialize $\nu^0$, arbitrarily.\\
For $k=1,2,\ldots$\\
$\;\;\;\;  (\gamma^k,\beta^k,\theta^k)=\arg\min_{(\gamma,\beta,\theta)} \mathcal{L}_c(\gamma,\beta,\theta,\mu^{k-1},\nu^{k-1})$\\
$\;\;\;\;\nu^{k}_i=\nu^{k-1}_i+c(\theta_i^k-\beta_i^k+\beta_{i-1}^k),i=2,\ldots,n$\\
$\;\;\;\;\mu^{k}_i=\mu^{k-1}_i+c(\gamma_i^k-\beta_i^k),i=1,\ldots,n$\\
\hline
\end{tabular}

\bigskip

\bigskip

\begin{tabular}{|l|}
\hline
Algorithm 4\\
\hline
initialize $\nu^0$, $\beta^0$ and $\theta^0$, arbitrarily.\\
For $k=1,2,\ldots$\\
$\;\;\;\;  \gamma^k=\arg\min_{\gamma}\mathcal{L}_c(\gamma,\beta^{k-1},\theta^{k-1},\mu^{k-1},\nu^{k-1})$\\
$\;\;\;\;  \beta^k=\arg\min_{\beta} \mathcal{L}_c(\gamma^{k},\beta,\theta^{k-1},\mu^{k-1},\nu^{k-1})$\\
$\;\;\;\;  \theta^k=\arg\min_{\theta} \mathcal{L}_c(\gamma^{k},\beta^{k},\theta,\mu^{k-1},\nu^{k-1})$\\
$\;\;\;\;  \nu^{k}_i=\nu^{k-1}_i+c(\theta_i^k-\beta_i^k+\beta_{i-1}^k),i=2,\ldots,n$\\
$\;\;\;\;\mu^{k}_i=\mu^{k-1}_i+c(\gamma_i^k-\beta_i^k),i=1,\ldots,n$\\
\hline
\end{tabular}

\bigskip

In Algorithm 3, $\arg\min_{(\gamma,\beta,\theta)} \mathcal{L}_c(\gamma,\beta,\theta,\mu^{k-1},\nu^{k-1})$ is typically difficult to find directly and iterative updating of each one of them with others fixed is applied (i.e., repeat the first three steps in the loop of Algorithm 4 until convergence). We will use simulation later to compare the relative efficiency of Algorithm 3 and Algorithm 4. 

We now consider each update in detail. Note the doubly augmented Lagrangian is 
\begin{eqnarray*}
\mathcal{L}_c(\gamma,\beta,\theta,\mu,\nu)&=&\sum_{i=1}^nF_i(y_i,\gamma_i)+\lambda_1\sum_{i=1}^n|\gamma_i|+\lambda_2\sum_{i=2}^n|\theta_i|+\sum_{i=1}^n\mu_i(\gamma_i-\beta_i)+\frac{c}{2}\sum_{i=1}^n(\gamma_i-\beta_i)^2\\
&&+\sum_{i=2}^n\nu_i(\theta_i-\beta_i+\beta_{i-1})+\frac{c}{2}\sum_{i=2}^n(\theta_i-\beta_i+\beta_{i-1})^2.
\end{eqnarray*}
The update for $\beta$ can be performed in closed form by solving a linear system, which still involves a tridiagonal matrix and can be solved efficiently. Note that the effect of introducing $\gamma_i$ is to decouple some terms in the Lagrangian so that the loss function and the lasso penalty do not come into play when updating $\beta$. The update for $\theta$ is the same as before and can be performed with component-wise thresholding using the same formula (\ref{eqn:updatetheta}). The update for $\gamma$ is generally not available in closed form. However, due to the special separable structure of the functional, it can be updated component by component, resulting in multiple one-dimensional convex optimization problems for which many efficient solvers exist. For different convex loss, only the updates for $\gamma_i$ need to be modified. We also note that for the quadratic loss, the updates for $\gamma_i$ is also a simple soft thresholding. 

\textit{Example. } In this example we take $F_i(y_i,\gamma_i)=|y_i-\gamma_i|$, the absolute deviation or $L_1$ loss. The $L_1$ loss function is an interesting alternative to the quadratic loss in that it is more robust to outliers. We refer to the resulting FLSA problem (\ref{eqn:flsa}) with $L_1$ loss as LAD-FLASSO. In this case, the update of $\gamma_i$ consists in minimizing $|y_i-\gamma_i|+\lambda_1|\gamma_i|+\mu_i(\gamma_i-\beta_i)+c/2(\gamma_i-\beta_i)^2$. Although the solution is not available in closed form, the function is strictly convex and differentiable except at two points, $0$ and $y_i$. Thus the minimizer can be found by comparing its values at $0$, $y_i$, and other potential stationary points, a total of only six cases (by considering the sign of $\gamma_i$ and $y_i-\gamma_i$). Thus the update in $\gamma$ can also be found efficiently and implemented easily. $\Box$

In the following theorem, we give the convergence of Algorithms 1-4. It shows that $\beta^k$ is a minimizing sequence of the FLSA (\ref{eqn:flsa}). If the minimizer is unique, then $\beta^k$ converges to the minimizer. The proof of the theorem is given in Appendix A in the Supplementary Material.

\begin{thm}For any of the algorithms 1-4, we have $f(\beta^k)\rightarrow \min_\beta f(\beta)$ where $f$ is the FLSA functional defined in (\ref{eqn:flsa}).
\end{thm}

\section{Simulation Results}
We follow the similar simulation setups used in \cite{hoefling10}. Each simulated sequence consists of data points with values of 0, 1, 2 and roughly 20\% of the data points have value 1 and another 20\% have value 2, with Gaussian noises added (except in Experiment 4 below where noise with heavy-tailed distribution is used). In experiments 1-3 below, we restrict ourselves to quadratic loss functions. The experiments are performed on HP workstation xw4400 with Intel Core 2 Duo Processor 2.66GHz and 2GB of RAM, implemented in R. We also make use of the limSolve package in R which implemented the tridiagonal matrix algorithm. We apply our doubly augmented Lagrangian method to the simulated dataset. The different between algorithm 3 and algorithm 4 is that algorithm 3 has an additional inner loop that applies the first three updatings in the loop of Algorithm 4 repeatedly till convergence. 

\textit{Experiment 1.} First we study the effect of the number of iterations, $T$, performed in this inner loop. Thus Algorithm 3 corresponds to the case $T\rightarrow\infty$ while Algorithm 4 corresponds to $T=1$. In this experiment, we set the sequence length $n=200$ and $n=2000$, with Gaussian noise of variance $0.1$, $c=5$ and $\lambda_1=0.5, \lambda_2=4$. 100 datasets are simulated in this experiment. The convergence criterion used is $||\nu^{k}-\nu^{k-1}||+||\mu^{k}-\mu^{k-1}||<10^{-10}$. In Table \ref{tab:T}, we show the average number of iterations required till convergence as well as the time (in seconds) elapsed. We see that although using $T>1$ reduced the number of iterations (for the outer loop) required, the overall computation time is either similar to the case with $T=1$ or significantly increased even for small value of $T$. Thus we see no advantage of using $T>1$ and Algorithm 4 is adopted in the following. We have also conducted simulations using other sequence lengths and parameters and the conclusion is the same.

\begin{table}
\caption{Simulation results for Experiment 1 for varying the number of iterations of the inner loop.\label{tab:T}}
\begin{tabular}{ccccccccc}
\hline
       &\multicolumn{4}{c}{n=200}&\multicolumn{4}{c}{n=2000}\\
\hline
               &T=1&T=2&T=5&T=10 &T=1&T=2&T=5&T=10\\
\hline 
number of iterations & 226.95 &131.61&  72.70 & 69.09&  212.02& 147.84 & 78.56  &71.21\\
computation time   &0.117 & 0.118& 0.144& 0.258& 0.354& 0.654& 0.838&1.753\\
\hline
\end{tabular}
\end{table}

\begin{figure}
\centerline{
\subfigure[]{\includegraphics[width=1.9in]{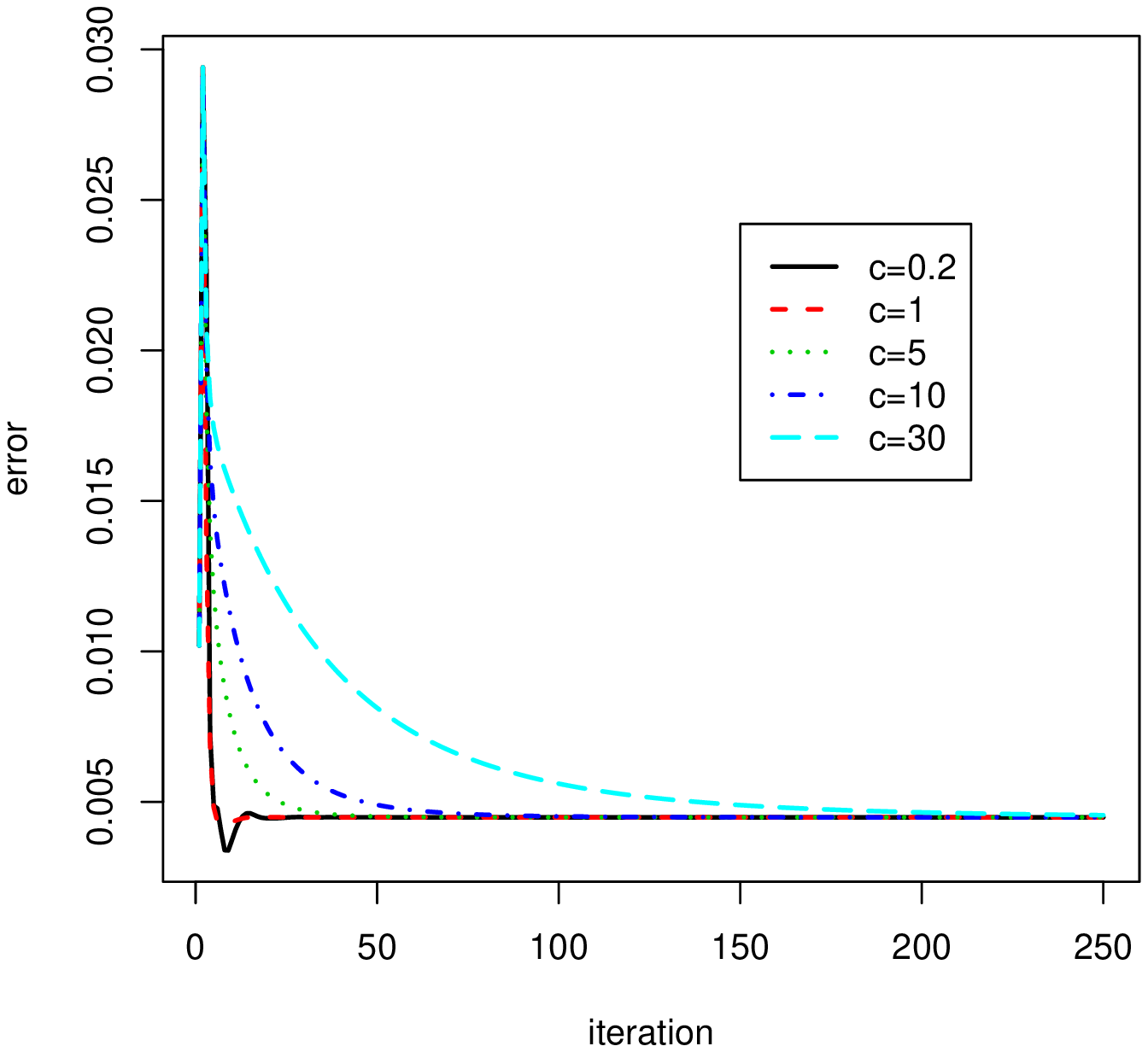}
}
\hfil
\subfigure[]{\includegraphics[width=1.9in]{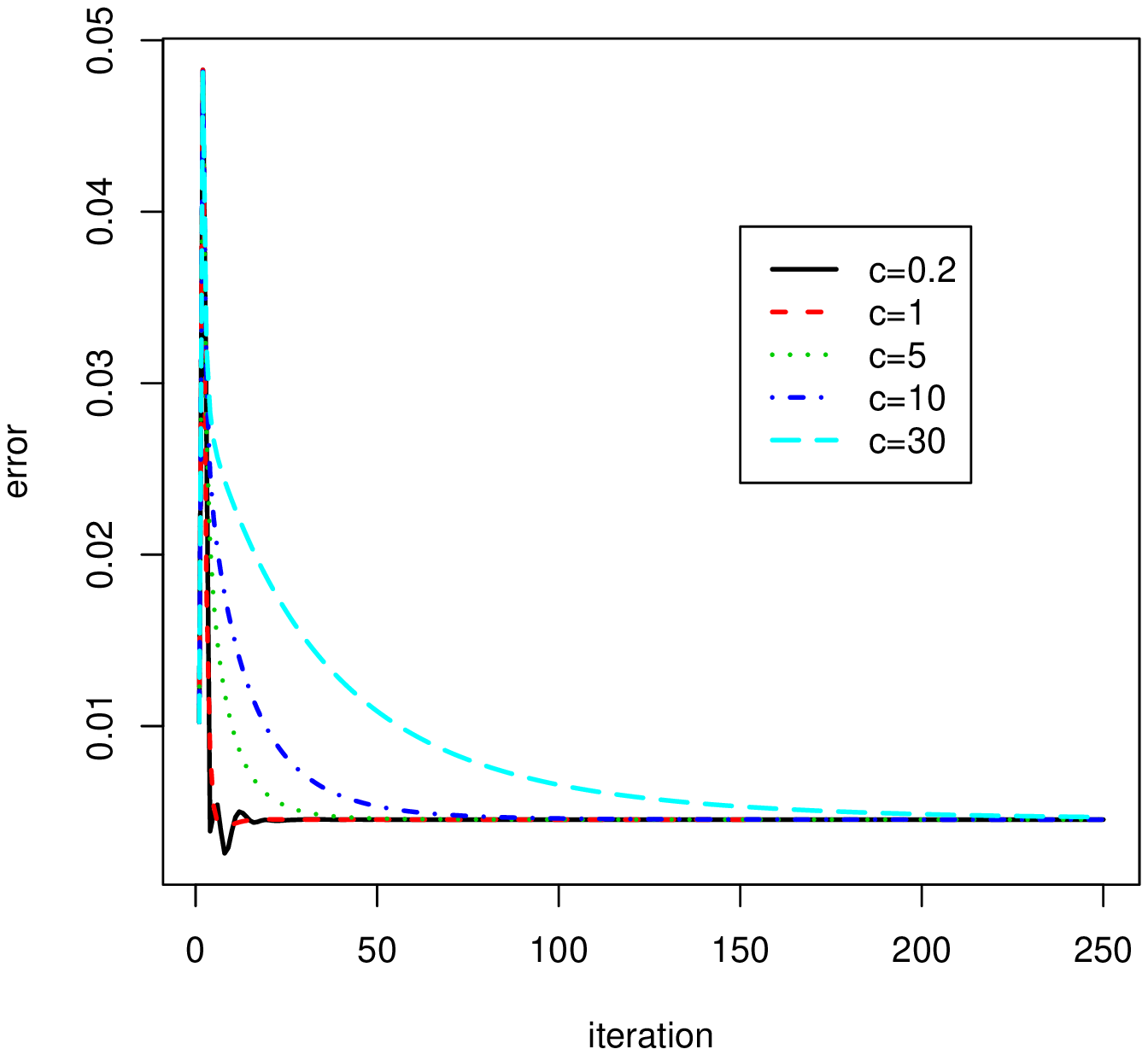}
}
\hfil
\subfigure[]{\includegraphics[width=1.9in]{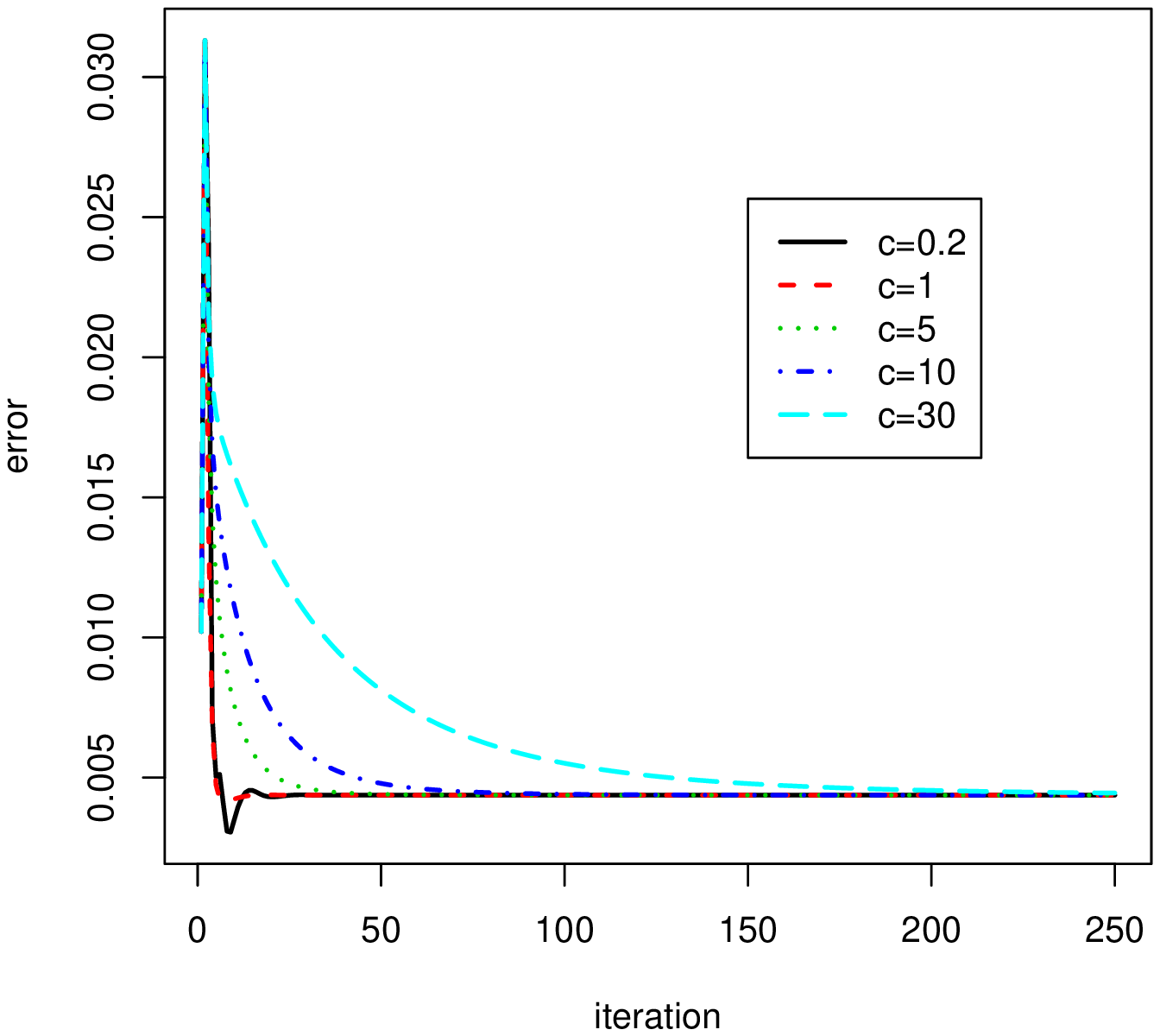}
}}

\centerline{\subfigure[]{\includegraphics[width=1.9in]{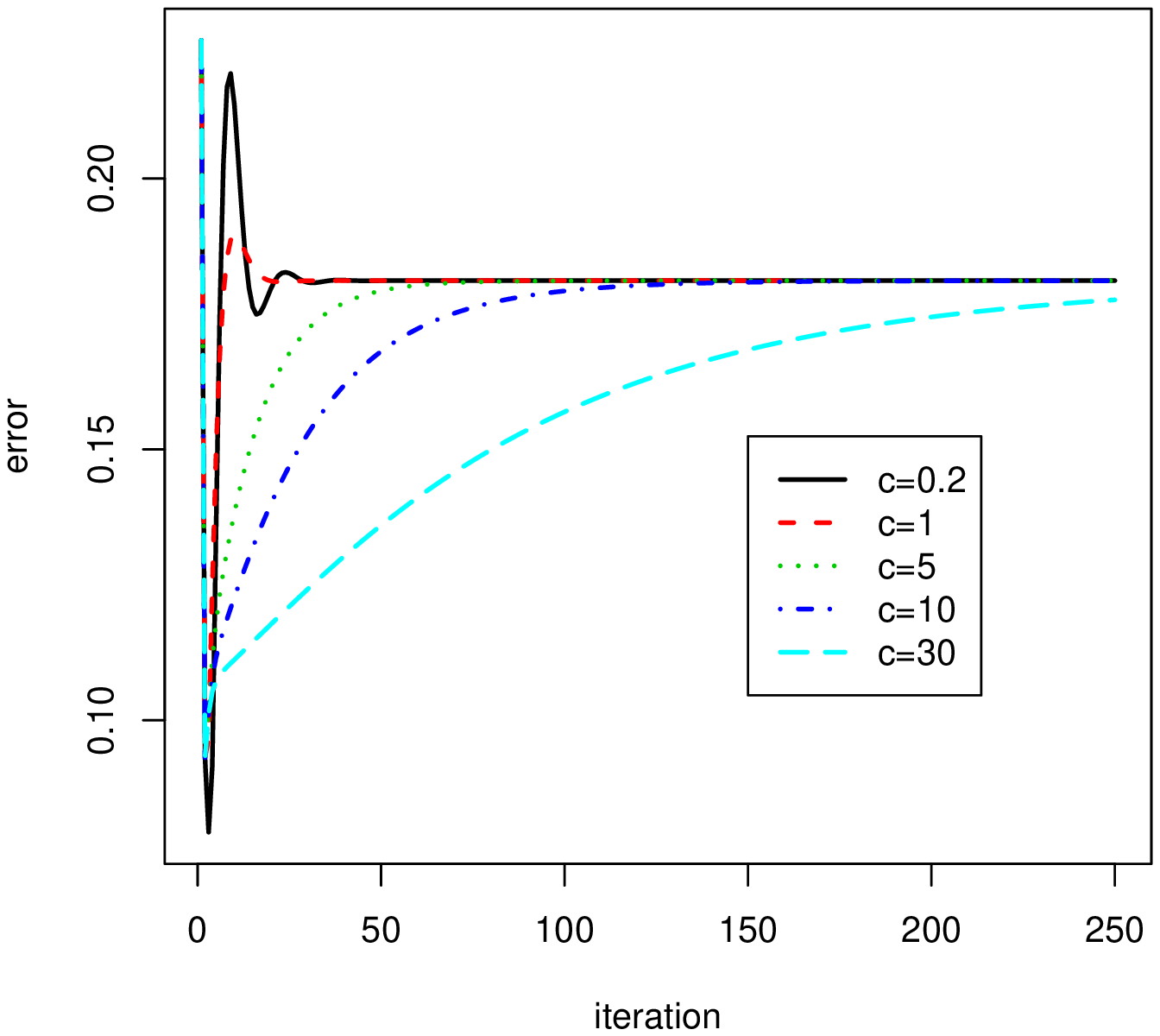}
}
\hfil
\subfigure[]{\includegraphics[width=1.9in]{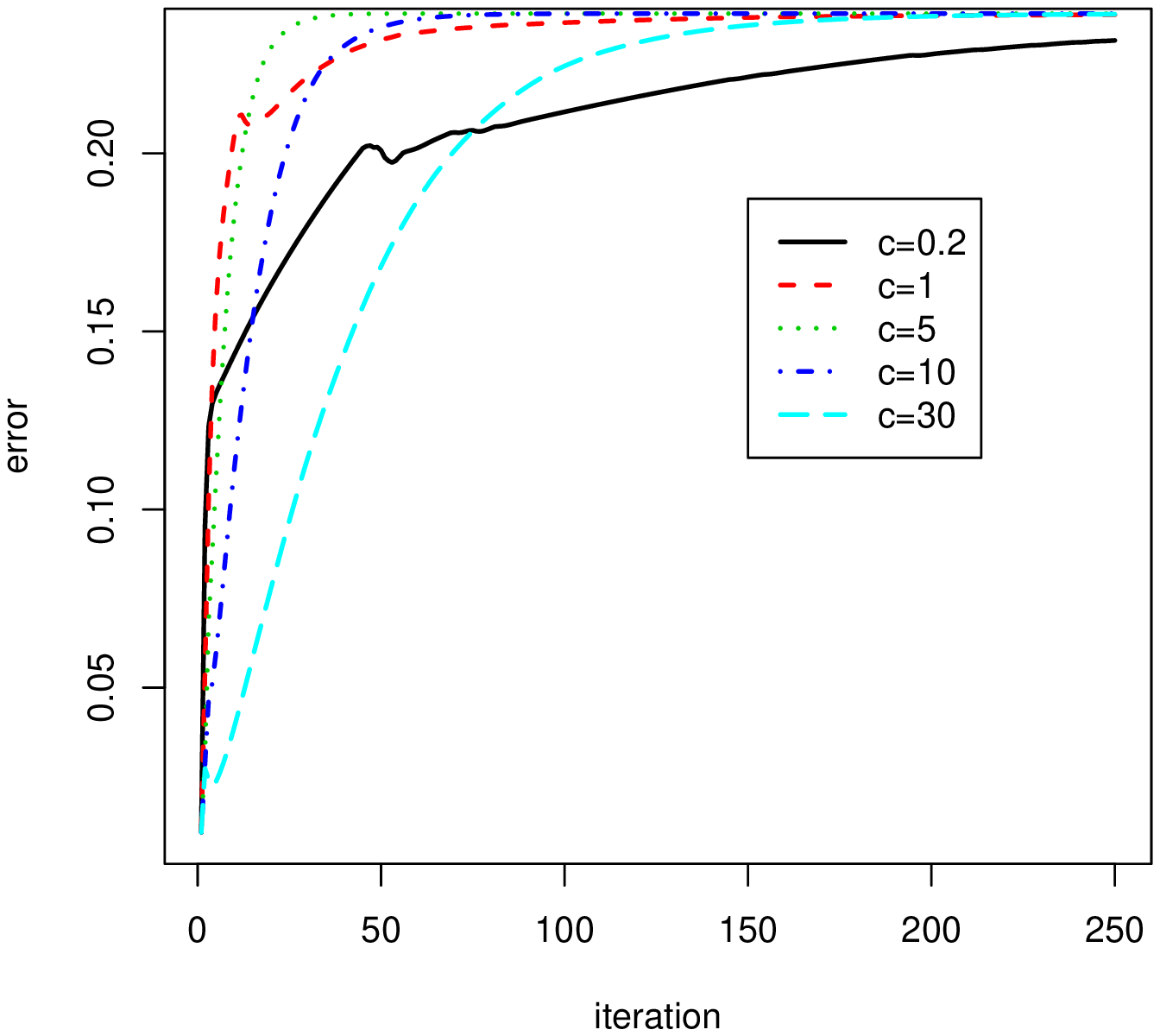}
}}

\centerline{\subfigure[]{\includegraphics[width=1.9in]{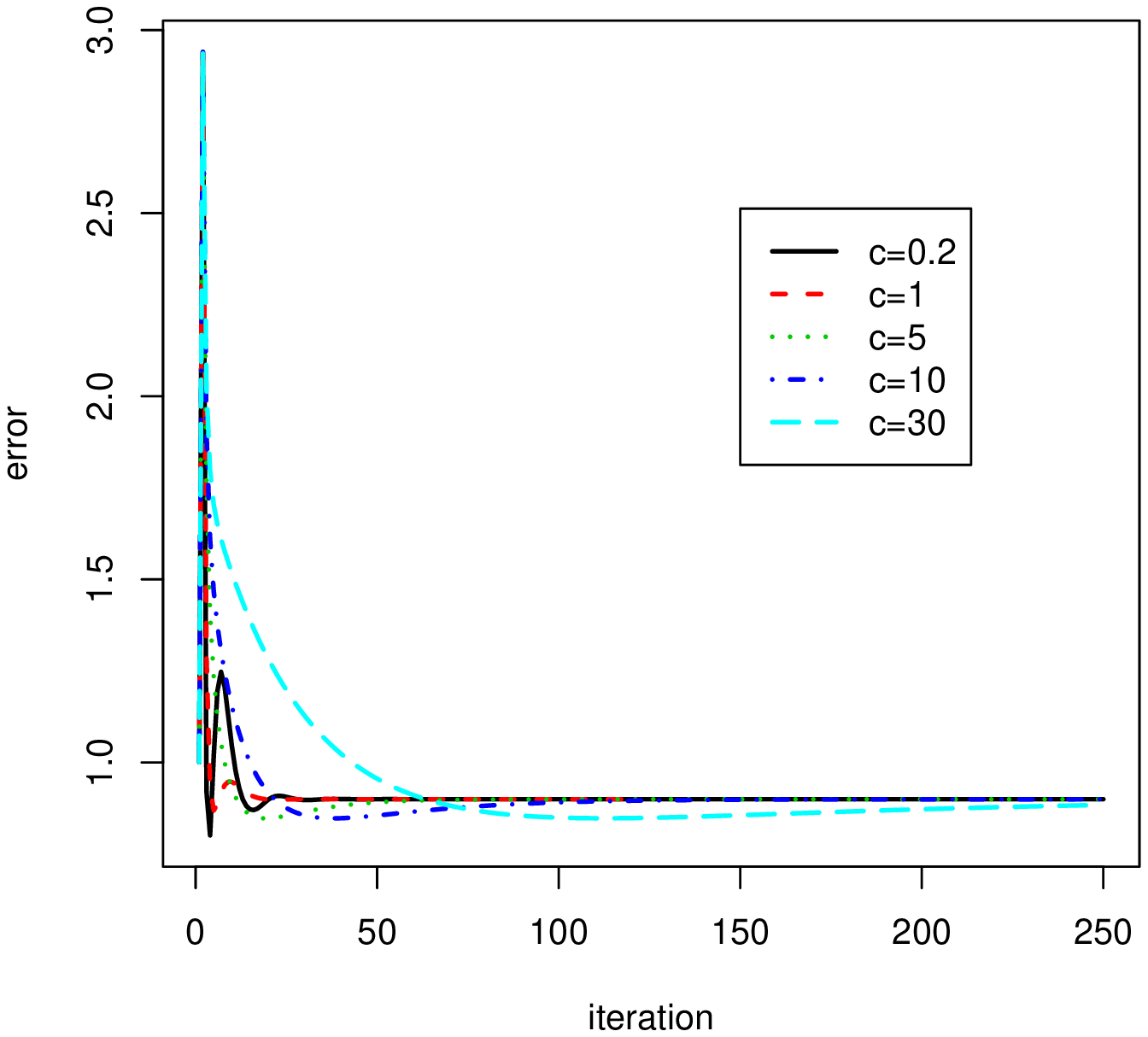}
}
\hfil
\subfigure[]{\includegraphics[width=1.9in]{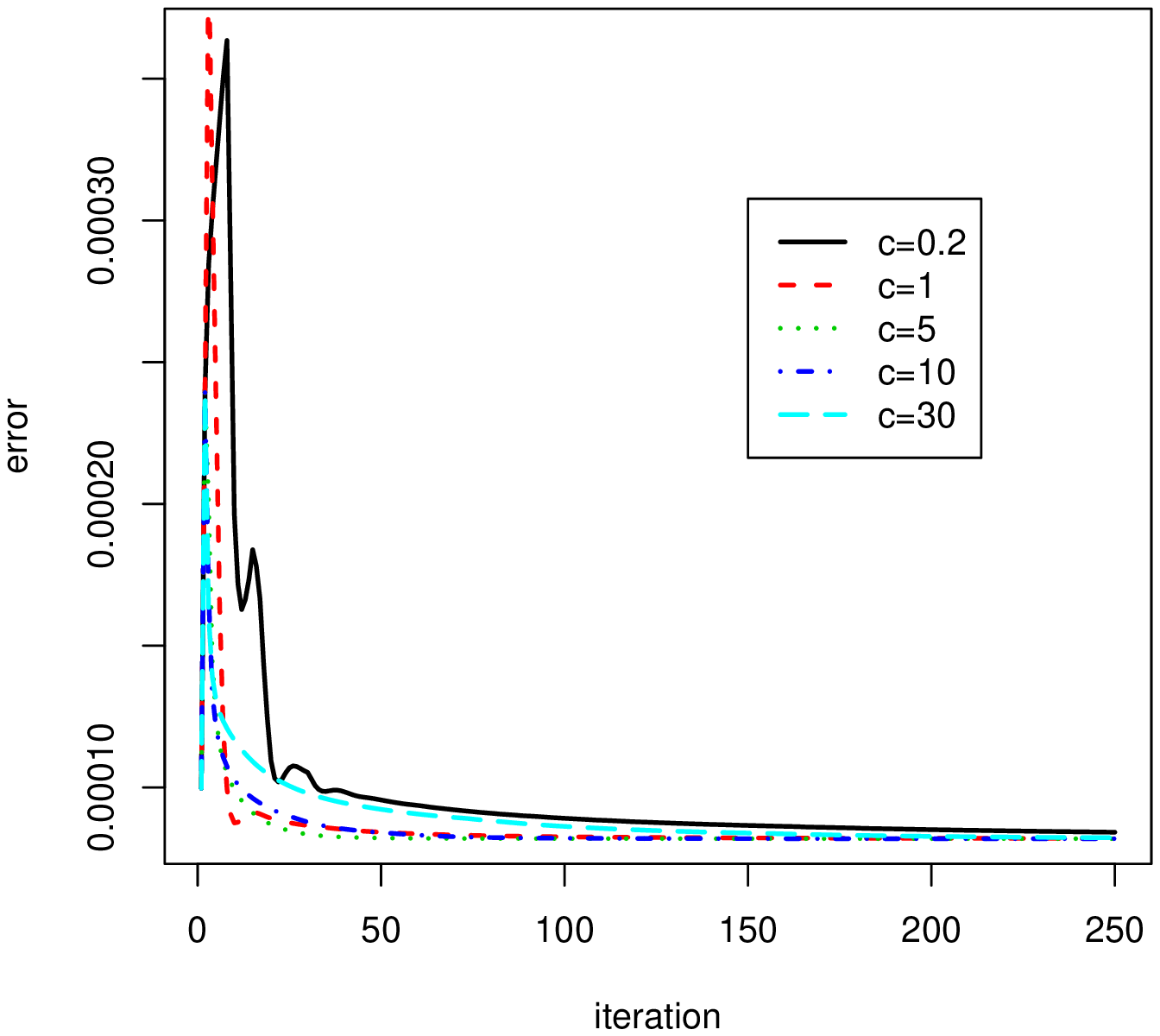}
}}
\caption{Evolution of mean squared error of reconstructed signals with iterations. \label{fig:c}   }
\end{figure}

\textit{Experiment 2.} Next we consider the effect of the parameter $c$ on the convergence of the algorithm. Although theoretically the ALM converges in the limit for any $c>0$, we will see that this parameter can affect the speed of convergence. Our simulation involves a sequence of length $n=1000$ with $N(0,0.1)$ noises, and we solve the FLSA problem with $\lambda_1=0$ and $\lambda_2=0.1$. We choose many different values for $c$ and the evolution of the mean squared error $\sum_{i=1}^n(\beta_i^k-\beta_i)^2/n$ for five values of $c$ is plotted in Figure \ref{fig:c} (a). Here $\beta_i$ represents the true signal and $\beta_i^k$ is the estimate for the $k$-th iteration. We see that for small value of $c$, the convergence of the estimate is slow and oscillate in the initial stage, while for big values of $c$, the convergence is also slow. For this sequence, a value between $0.5$ and $5$ generally produces reasonable speed of convergence visually. 

Now we vary different parameters involved in the optimization to investigate how these changes affect the choice of $c$. First we generate a sequence of length $n=100$ and another with $n=10000$. The convergence diagnostic plots are shown in Figure \ref{fig:c} (b) and (c). Remarkably, the plots show that the choice of $c$ is almost unaffected by the length of the sequence and the number of iterations required for convergence does not depend on $n$ . This empirical observation has at least two implications. (i) In order to choose a reasonable value of $c$ for an extremely long sequence, we can run the algorithm on a subsequence with several different values of $c$ and choose the best one based on the convergence speed on the subsequence. Of course for this to work we need to assume the sequence is stationary in some sense. (ii) The complexity of the algorithm is linear in the length of the sequence since it is linear for each iteration and the number of iterations does not vary much with the length (note this is only based on empirical observation). 

Then we use the same sequence with $n=1000$ but with a bigger noise variance $0.4$. The plot shown in Figure \ref{fig:c} (d) looks different, but the range of values for $c$ that results in fast convergence is similar as before. In  Figure \ref{fig:c} (e), we show the results when solving FLSA with $\lambda_1=0.5$ and $\lambda_2=4$, and in Figure \ref{fig:c} (f),(g), we multiply and divide the noisy sequence by a factor of $10$ respectively. When these parameters are changed, we see the trace plot is more variable. Since all these types of changes can be regarded as the change in relative sizes of the different terms in (\ref{eqn:lagrangian2}), we conclude that the choice of $c$ depends on this relative scale but is quite stable otherwise. Even so, we still observe that the optimal choice of $c$ is somewhere between $0.2$ and $10$. We have generated different sequences and worked with different regularization parameters to make sure the observations made above apply to a wide variety of settings. Since our algorithm is relatively fast, we can suggest running the algorithm for several different values of $c$ and visually check its convergence, except when the sequence is extremely long ($>10^6$) and then we can run the algorithm on one or more subsequences to choose $c$ before running it on the entire sequence. In all the following experiments we set $c=5$.

\textit{Experiment 3.} Here we want to say something about the computation speed of our algorithm based on comparisons with previous approaches. We download from CRAN the flsa package (version 1.03) which is based on the path algorithm presented in \cite{hoefling10}. For each case of $n=100, 1000, 10000,100000$, we generate $100$ sequences and the average computation times for each sequence are shown in Table \ref{tab:flsa} for $\lambda_1=0.5, \lambda_2=4$. From the results reported in the table, we see that the path algorithm is about 20 times faster than our ALM algorithm for $n\ge 1000$. For the case $n=100$, the difference is about 100 fold. 
We observe from Table \ref{tab:flsa} that both algorithms have computation time approximately linear in $n$, except for ALM when $n=100$. Thus the large difference for $n=100$ may be due to the reason that in this case most of the computation time in ALM is spent on ancillary chores such as calling the R function, setting up parameter values and returning results. The reported computation time for the path algorithm is slower than those reported in \cite{hoefling10} and the reason might be due to the difference in simulation setup and difference in computer system used.

We also need to note that the path algorithm is specifically designed for computing the entire solution path for all regularization parameters, and in this sense it should have even better performance when the solution for many regularization parameter values are sought. However, this algorithm does not work with general loss function as the ALM does. 

There is no publicly available package implementing the descent algorithm in \cite{friedman07}. However, Table 1 in \cite{hoefling10} reported that the path algorithm is about 10-100 times faster than the descent algorithm when $n\le 10^4$, while the two algorithms have comparable speed with larger $n$. Based on this comparison, we think our algorithm is probably comparable to the descent algorithm when $n\le 10^4$ but much slower for bigger sequence length.  Finally, we note it is difficult to exactly compare the computation speed for different algorithms since all algorithms involve some parameter choice. In particular, the convergence criterion used in our implementation is $||\nu^k-\nu^{k-1}||+||\mu^k-\mu^{k-1}||\le 10^{-10}$, and if we increase the threshold to $10^{-5}$, it becomes 3 to 5 times faster. Besides, the flsa package uses C code in its underlying implementation which makes it faster, and our implementation uses tridiagonal matrix algorithm from the limSolve package which uses Fortran code in its implementation and thus the net effect is difficult to compare. We emphasize again that the biggest advantage of our algorithm is the ease in implementation as well as that it works with general convex loss functions. 

\begin{table}
\caption{Comparison of computation speed for our ALM algorithm and the flsa package based on the path algorithm.\label{tab:flsa}}
\begin{tabular}{cccccc}
\hline
       \\
\hline
       &$n=100$&$n=1000$&$n=10^4$&$n=10^5$&$n=10^6$\\
\hline 
ALM &0.09811&0.1797&1.861&21.702&223.9\\
flsa &0.00092&0.0073&0.072&0.958&10.51\\
\hline
\end{tabular}
\end{table}

\begin{figure}
\centering
\includegraphics[width=5in]{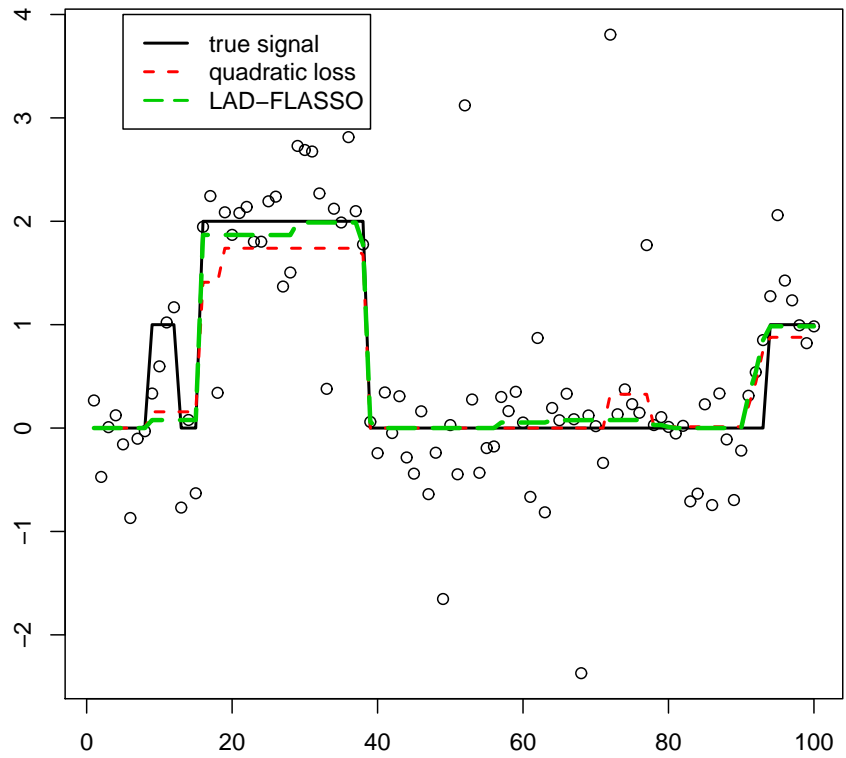}

\caption{Reconstruction of a signal sequence based on FLSA with quadratic loss and absolute deviation ($L_1$) loss.\label{fig:lad}   }
\end{figure}

\textit{Experiment 4.} Finally, in this experiment, we consider the LAD-FLASSO problem, where the loss function in (\ref{eqn:flsa}) is defined by $F_i(y_i,\beta_i)=|y_i-\beta_i|$. We only illustrate here with a sequence of length $n=100$ and the noise has t distribution with 2 degrees of freedom and the scale parameter equal to $0.3$. With heavy-tailed noises, the LAD-FLASSO is expected to perform better than the usual FLSA with quadratic loss. Indeed, Figure \ref{fig:lad} shows the noisy sequence, the true signal, as well as the two reconstructions. The regularization parameters $\lambda_1$ and $\lambda_2$ in the two cases are those minimizing sum of squared errors and sum of absolute deviations respectively (of course this depends on the knowledge of the true signal in the simulation), by searching over a fine grid. An obvious difference between the two reconstructions is seen at positions 70-80, where an extremely high value of observation occurs due to the heavy-tailed noise distribution. 

\section{Conclusion}
In this paper we propose a simple algorithm for the FLSA problem. Although not as fast as the path algorithm implemented in the flsa package, the most attractive feature of this algorithm is the simplicity of its implementation and it works for general convex loss functions. However, the computational speed of the current implementation in R can possibly be improved if a more general programming language such as $C$ is used for its underlying implementation. Another advantage of the algorithm is that it is provably convergent for any initialization values, and its convergence properties are investigated based on simulation studies presented here. The flexibility in implementation is demonstrated by our implementation of the LAD-FLASSO problem which is lacking from other existing implementations based on either descent algorithm or path algorithm. We expect that ALM as a general technique will be very useful for computing other optimization problems in statistical learning. 

\bibliographystyle{jasa}
\bibliography{papers.txt,books.txt}

\section*{Supplementary Material}
\section*{Appendix A Proof of Theorem 1.} In the proof we use matrix and vector notations. In particular, the expressions $\theta_i-\beta_i+\beta_{i-1}, i=2,\ldots, n $ can be written as $\theta-A\beta$ with $A$ an $(n-1)\times n$ matrix. We also make frequent use of some standard and classical results from convex analysis, such as those contained in \cite{rockafellar70,ekeland83}, most notably the properties of subdifferential for convex functions. Also, we only show the convergence of Algorithms 1 and 2 while the analysis for Algorithms 3 and 4 is very much the same but more tedious to write down and thus omitted.

We start with Algorithm 1, for which the augmented Lagrangian can be written as
\begin{equation*}
\mathcal{L}(\beta,\theta,\nu)=U(\beta)+V(\theta)+\frac{c}{2}||\theta-A\beta||^2+\nu^T(\theta-A\beta),
\end{equation*}
where $U(\beta)=\sum_{i=1}^nF_i(y_i,\beta_i)+\lambda_1\sum_{i=1}^n|\beta_i|$, $V(\theta)=\lambda_2\sum_{i=2}^n|\theta_i|$, and $\nu^T$ is the transpose of the column vector $\nu$. In the proof we only need to use the convexity of $U$ and $V$. 

Using the usual notation, suppose $(\beta^*,\theta^*,\nu^*)$ is the saddle point of $\mathcal{L}$ satisfying 
\begin{equation}\label{eqn:lagrangian}
\mathcal{L}(\beta^*,\theta^*,\nu)\le \mathcal{L}(\beta^*,\theta^*,\nu^*)  \le \mathcal{L}(\beta,\theta,\nu^*) \; \forall \beta,\theta,\nu
\end{equation}

From the first equality of (\ref{eqn:lagrangian}), we have $\theta^*=A\beta^*$. The update for $\nu$ in Algorithm 1 is 
$\nu^{k}=\nu^{k-1}+c(\theta^k-A\beta^k)$, which implies 
\begin{equation}\label{eqn:updatenu}
\bar{\nu}^{k}=\bar{\nu}^{k-1}+c(\bar{\theta}^k-A\bar{\beta}^k),
\end{equation}
where we set $\bar{\beta}^k=\beta^k-\beta^*, \bar{\theta}^k=\theta^k-\theta^*$ and $\bar{\nu}^{k}=\nu^{k}-\nu^*$.
From (\ref{eqn:updatenu}), we immediately get
\begin{equation*}
||\bar{\nu}^{k-1}||^2-||\bar{\nu}^{k}||^2=-2c(\bar{\nu}^{k-1})^T(\bar{\theta}^k-A\bar{\beta}^k)-c^2||\bar{\theta}^k-A\bar{\beta}^k||^2.
\end{equation*}
Next we show the right hand side of the above is nonnegative.

 From the second inequality of $(\ref{eqn:lagrangian})$, we have
\begin{eqnarray}
0\in\partial_\beta\mathcal{L}(\beta^*,\theta^*,\nu^*)&\Leftrightarrow&0\in\partial U(\beta^*)-cA^T(\theta^*-A\beta^*)-{\nu^*}^TA,\label{eqn:1}\\
0\in\partial_\theta\mathcal{L}(\beta^*,\theta^*,\nu^*)&\Leftrightarrow&0\in\partial V(\theta^*)+c(\theta^*-A\beta^*)+{\nu^*},\label{eqn:2}
\end{eqnarray}
where $\partial$ is the notation for the subdifferential of a convex function. 

Correspondingly, based on the update of $\beta^k$ and $\theta^k$ in Algorithm 1, we have
\begin{eqnarray}
0\in\partial_\beta\mathcal{L}(\beta^k,\theta^k,\nu^{k-1})&\Leftrightarrow&0\in\partial U(\beta^k)-cA^T(\theta^k-A\beta^k)-(\nu^{k-1})^TA,\label{eqn:3}\\
0\in\partial_\theta\mathcal{L}(\beta^k,\theta^k,\nu^{k-1})&\Leftrightarrow&0\in\partial V(\theta^k)+c(\theta^k-A\beta^k)+{\nu^{k-1}}.\label{eqn:4}
\end{eqnarray}

Subtracting (\ref{eqn:1}) from (\ref{eqn:3}) and subtracting (\ref{eqn:2}) from (\ref{eqn:4}), we get
\begin{eqnarray}
0&\in&\partial U(\beta^k)-\partial U(\beta^*)-cA^T(\bar{\theta}^k-A\bar{\beta}^k)-({\bar{\nu}}^{k-1})^TA,\label{eqn:U}\\
0&\in&\partial V(\theta^k)-\partial V(\theta^*)+c(\bar{\theta}^k-A\bar{\beta}^k)+\bar{\nu}^{k-1}.\label{eqn:V}
\end{eqnarray}

Multiplying $(\bar{\beta}^k)^T$ to (\ref{eqn:U}) from the left, multiplying $(\bar{\theta}^k)^T$ to (\ref{eqn:V}) from the left, and adding the two expressions gives us
\begin{equation}\label{eqn:add}
0\in\langle \partial U(\beta^k)-\partial U(\beta^*),\bar{\beta}^k\rangle+\langle \partial V(\theta^k)-\partial V(\theta^*),\bar{\theta}^k\rangle+c||\bar{\theta}^k-A\bar{\beta}^k||^2+(\bar{\nu}^{k-1})^T(\bar{\theta}^k-A\bar{\beta}^k),
\end{equation}
where we used $\langle \cdot,\cdot\rangle$ to denote the dot product of two vectors in some places above to be consistent with the usual notation in convex analysis as in \cite{ekeland83}.

From standard results in convex analysis, all elements in $\langle \partial U(\beta^k)-\partial U(\beta^*),\bar{\beta}^k\rangle$ and $\langle \partial V(\theta^k)-\partial V(\theta^*),\bar{\theta}^k\rangle$ are nonnegative and thus we get $c||\bar{\theta}^k-A\bar{\beta}^k||^2+(\bar{\nu}^{k-1})^T(\bar{\theta}^k-A\bar{\beta}^k)\le 0$ which immediately implies that
\begin{equation*}
||\bar{\nu}^{k-1}||^2-||\bar{\nu}^{k}||^2=-2c(\bar{\nu}^{k-1})^T(\bar{\theta}^k-A\bar{\beta}^k)-c^2||\bar{\theta}^k-A\bar{\beta}^k||^2\ge c^2||\bar{\theta}^k-A\bar{\beta}^k||^2.
\end{equation*}

Now that $||\bar{\nu}^k||^2$ is nonnegative and decreasing, we obtain $\bar{\theta}^k-A\bar{\beta}^k\rightarrow 0$. Using this in (\ref{eqn:add}), we get 
\begin{equation}\label{eqn:convzero}
0\le \langle\partial U(\beta^k)-\partial U(\beta^*),\bar{\beta}^k\rangle\rightarrow 0,\;0\le\langle \partial V(\theta^k)-\partial V(\theta^*),\bar{\theta}^k\rangle\rightarrow 0,
\end{equation}
where the above expression is taken to mean that ``there exists some sequence $u_k\in \langle\partial U(\beta^k)-\partial U(\beta^*),\bar{\beta}^k\rangle$ with $0\le u_k\rightarrow 0$", for example. Similar interpretations are used in the following.

By the definition of subdifferential, we have
\begin{eqnarray*}
U(\beta^k)&\ge& U(\beta^*)+\langle\partial U(\beta^*),\bar{\beta}^k\rangle,\\
U(\beta^*)&\ge& U(\beta^k)-\langle\partial U(\beta^k),\bar{\beta}^k\rangle,\\
\end{eqnarray*}
resulting in 
\begin{eqnarray*}
U(\beta^k)-\langle\partial U(\beta^*),\bar{\beta}^k\rangle&\ge& U(\beta^*)\ge U(\beta^k)-\langle\partial U(\beta^k),\bar{\beta}^k\rangle.
\end{eqnarray*}
Using (\ref{eqn:convzero}), the difference between and left hand side and the right hand side is converging to zero and thus we have $U(\beta^k)\rightarrow U(\beta^*)$. Similarly we can show $V(\theta^k)\rightarrow V(\theta^*)$. These combined with $\bar{\theta}^k-A\bar{\beta}^k\rightarrow 0$ prove the convergence of Algorithm 1.

For Algorithm 2, the proof strategy is similar and we only point out the differences. The proof is the same as before up to equation (\ref{eqn:2}). Because the order of update of $\beta$ and $\theta$ in Algorithm 2, equation (\ref{eqn:3}) is replaced by
\begin{eqnarray*}
0\in\partial_\beta\mathcal{L}(\beta^k,\theta^{k-1},\nu^{k-1})&\Leftrightarrow&0\in\partial U(\beta^k)-cA^T(\theta^{k-1}-A\beta^k)-(\nu^{k-1})^TA,
\end{eqnarray*}
and thus equation (\ref{eqn:U}) becomes instead 
\begin{eqnarray*}
0&\in&\partial U(\beta^k)-\partial U(\beta^*)-cA^T(\bar{\theta}^{k-1}-A\bar{\beta}^k)-({\bar{\nu}}^{k-1})^TA,
\end{eqnarray*}
while equation (\ref{eqn:V}) remains the same. Then we have, in place of (\ref{eqn:add}),
\begin{eqnarray*}
0&\in&\langle \partial U(\beta^k)-\partial U(\beta^*),\bar{\beta}^k\rangle+\langle \partial V(\theta^k)-\partial V(\theta^*),\bar{\theta}^k\rangle\\
 &&+c||\bar{\theta}^k-A\bar{\beta}^k||^2+(\bar{\nu}^{k-1})^T(\bar{\theta}^k-A\bar{\beta}^k)-c(\bar{\beta}^k)^TA^T(\bar{\theta}^{k-1}-\bar{\theta}^k),
\end{eqnarray*}
which then implies 
\begin{equation}\label{eqn:nuconv}
||\bar{\nu}^{k-1}||^2-||\bar{\nu}^{k}||^2\ge c^2||\bar{\theta}^k-A\bar{\beta}^k||^2-2c^2(\bar{\beta}^k)^TA^T(\bar{\theta}^{k-1}-\bar{\theta}^k).
\end{equation}
So the difference from the corresponding analysis for Algorithm 1 is the extra term $-2c^2(\bar{\beta}^k)^TA^T(\bar{\theta}^{k-1}-\bar{\theta}^k)$ on the right hand side above.

Now we analyze the term $2c^2(\bar{\beta}^k)^TA^T(\bar{\theta}^{k}-\bar{\theta}^{k-1})$. From (\ref{eqn:V}) (which is still true for Algorithm 2) and the update rule for $\nu$ in Algorithm 2, we have
\begin{eqnarray}
0&\in&\partial V(\theta^k)-\partial V(\theta^*)+c(\bar{\theta}^k-A\bar{\beta}^k)+\bar{\nu}^{k-1}\label{eqn:21}\\
0&\in&\partial V(\theta^{k-1})-\partial V(\theta^*)+c(\bar{\theta}^{k-1}-A\bar{\beta}^{k-1})+\bar{\nu}^{k-2}\label{eqn:22}\\
\bar{\nu}^{k-1}-\bar{\nu}^{k-2}&=&c(\bar{\theta}^{k-1}-A\bar{\beta}^{k-1}).\label{eqn:23}
\end{eqnarray} 
Subtracting (\ref{eqn:22}) from (\ref{eqn:21}) and taking into account (\ref{eqn:23}), we get
\begin{equation*}
0\in \partial V(\theta^k)-\partial V(\theta^{k-1})+c(\bar{\theta}^{k}-A\bar{\beta}^{k}).
\end{equation*}
Taking inner product with $\theta^k-\theta^{k-1}$ in the above equation and using the property of convex function that $\langle \partial V(\theta^k)-\partial V(\theta^{k-1}), \theta^k-\theta^{k-1}\rangle\ge 0$, we get 
\begin{equation*}
(\bar{\theta}^k-A\bar{\beta}^k)^T(\theta^k-\theta^{k-1})\le 0,
\end{equation*}
and we can rewrite the above expression as
\begin{equation*}
(\bar{\beta}^k)^TA^T(\bar{\theta}^k-\bar{\theta}^{k-1})\ge (\bar{\theta}^k)^T(\bar{\theta}^k-\bar{\theta}^{k-1}).
\end{equation*}
Using the identity $(\bar{\theta}^k)^T(\bar{\theta}^k-\bar{\theta}^{k-1})=1/2 (||\bar{\theta}^k||^2-||\bar{\theta}^{k-1}||^2+||\bar{\theta}^k-\bar{\theta}^{k-1}||^2)$ we obtain from (\ref{eqn:nuconv})
\begin{equation*}
||\bar{\nu}^{k-1}||^2-||\bar{\nu}^{k}||^2\ge c^2||\bar{\theta}^k-A\bar{\beta}^k||^2+c^2(||\bar{\theta}^k||^2-||\bar{\theta}^{k-1}||^2+||\bar{\theta}^k-\bar{\theta}^{k-1}||^2).
\end{equation*}
After rearranging, we get 
\begin{equation*}
(||\bar{\nu}^{k-1}||^2+c^2||\bar{\theta}^{k-1}||^2)-(||\bar{\nu}^{k}||^2+c^2||\bar{\theta}^{k}||^2)\ge c^2||\bar{\theta}^k-A\bar{\beta}^k||^2+c^2||\bar{\theta}^k-\bar{\theta}^{k-1}||^2,
\end{equation*}
and then $||\bar{\theta}^k-A\bar{\beta}^k||\rightarrow 0$ and $||\bar{\theta}^k-\bar{\theta}^{k-1}||\rightarrow 0$. Now the rest of the analysis follows that for Algorithm 1 with no changes.

\section*{Appendix B R code for FLSA with quadratic loss}
\begin{verbatim}
flasso.alm<-function(y,lambda1,lambda2,C=5,tol=1e-10){
  n<-length(y)

  #initialization
  beta<-y
  theta<-rep(0,n-1)
  gamma<-rep(0,n)
  mu<-rep(0,n)
  nu<-rep(0,n-1)

  conv<-100
  iter<-0

  while (conv>tol){

    temp<-(y+C*beta-mu/2)/(1+C)
    gamma<-abs(temp)-lambda1/(1+C)
    gamma<-pmax(0,gamma)*sign(temp)

    ##compute rhs of the linear system for solving beta
    temp1<--C*gamma; temp2<--mu; temp3<-c(theta[1],diff(theta),-theta[n-1])*C;
    temp4<-c(nu[1],diff(nu),-nu[n-1]); rhs<-temp1+temp2+temp3+temp4
    ##compute the three diagonals in the linear system
    diag1<-rep(-C/2,n-1); 
    diag2<-c(C/2,rep(C,n-2),C/2)+rep(C/2,n)
    ##call the tridiagonal matrix algorithm
    beta<-Solve.tridiag(diag1,diag2,diag1, -rhs/2)
 
    temp<-diff(beta)-nu/b
    theta<-abs(temp)-lambda2/b
    theta<-pmax(0,theta)*sign(temp)

    premu<-mu
    mu<-mu+C*(gamma-beta)
    prenu<-nu
    nu<-nu+C*(theta-diff(beta))

    conv<-mean(c((nu-prenu)^2,(mu-premu)^2)) #used to test convergence
    iter<-iter+1
  }#while loop end

#return the estimated signal and number of iterations performed
list(beta=beta,iter=iter) 
}
\end{verbatim}

\end{document}